\documentclass [11pt]{article}
\usepackage{latexsym}
\usepackage{amsfonts}
\usepackage{times}
\usepackage[margin=1.0in]{geometry}

\newtheorem{theorem}{Theorem}[section]
\newtheorem{observation}[theorem]{Observation}

\newtheorem{lemma}[theorem]{Lemma}
\newtheorem{definition}[theorem]{Definition}

\newenvironment{proof}{\noindent {\bf Proof:}\quad}
{\hfill$\Box$ \par \medskip}

\newlength{\boxwidth}
\def\bigo{\mathrm{O}}
\def\reals{\ensuremath{\mathbb{R}}}
\newcommand{\Lav}{L_{\rm avg}}
\newcommand{\Nash}{\ensuremath{{\cal N}}}
\newcommand{\assign}{\ensuremath{{\cal A}}}
\newcommand{\altA}{\overline{n}}

\newcommand{\hide}[1]{}

\begin{document}

\title{Utilitarian Resource Assignment\thanks{Submitted for
journal publication.
This work was partially supported by
the IST Program of the EU under contract numbers
IST-1999-14186 (ALCOM-FT)
and IST-1999-14036 (RAND-APX).}}

\author{Petra Berenbrink
\\
     School of Computing Science \\
     Simon Fraser University
\and
Leslie Ann Goldberg
\\
Department of Computer Science\\
University of Warwick
\and
Paul W.~Goldberg
\\
Department of Computer Science\\
University of Warwick
\and
Russell Martin
\\
Department of Computer Science\\
University of Warwick
}

\maketitle

\begin{abstract}
This paper studies a resource allocation problem introduced by
Koutsoupias and Papadimitriou. The scenario is modelled as a
multiple-player game in which each player selects one of a finite
number of known resources. The cost to the player is the total weight
of all players who choose that resource, multiplied by the ``delay''
of that resource. Recent papers have studied the Nash equilibria and
social optima of this game in terms of the $L_\infty$ cost metric, in
which the social cost is taken to be the maximum cost to any player.
We study the $L_1$~variant of this game, in which the social cost is
taken to be the sum of the costs to the individual players, rather
than the maximum of these costs.  We give bounds on the size of the
coordination ratio, which is the ratio between the social cost
incurred by selfish behavior and the optimal social cost; we also
study the algorithmic problem of finding optimal (lowest-cost)
assignments and Nash Equilibria. Additionally, we obtain bounds on the
ratio between alternative Nash equilibria for some special cases of
the problem.
\end{abstract}

\thispagestyle{empty}

\newpage
\setcounter{page}{1}
\section{Introduction}\label{sec:intro}

This paper studies the resource allocation problem introduced by
Koutsoupias and Papadimitriou~\cite{kp}. In this problem, we are given
a collection of resources such as computer servers, printers, or
communication links, each of which is associated with a
``delay''\footnote{The delay is the reciprocal of the quantity
  commonly called the ``speed'' or ``capacity'' in related work. It is
  convenient to work in terms of the delay, as defined here, because
  this simplifies our results.}.  We are also given a collection of
tasks, each of which is associated with a ``weight'' corresponding to
its size.  Each task chooses a resource.  A given resource is shared
between its tasks in such a way that each of these tasks incurs a cost
corresponding to the time until the resource has completed its
work. For example, the task might model a routing request and the
resources might model parallel links of a network.  If routing
requests are broken into packets and these are sent in a round-robin
fashion, each request will finish at (approximately) the time that the
link finishes its work.

We assume that each task chooses its resource in a selfish manner,
minimizing its own cost.  Following~\cite{kp} we are interested in
determining the social cost of this selfish behavior.  Previous work
on this problem has measured ``social cost'' in terms of the
$L_\infty$ metric --- that is, the longest delay incurred by any task.
Our measure of social cost is the $L_1$ metric -- that is, the average
delay (over tasks). This is sometimes called the {\em utilitarian}
interpretation of social welfare, and is a standard assumption in the
multi-agent system literature, for example~\cite{emst, mpw,
  sandholm}. In many settings, the average delay may be a better
measure of the quality of a solution than the very worst delay. Thus,
the $L_1$ metric is quite natural. This metric was also used in the
model of~\cite{RT} in the setting of infinitely many tasks.

We give bounds on the size of the coordination ratio, which is the
ratio between the social cost incurred by selfish behavior and the
optimal social cost~\cite{kp}; we also study the algorithmic problem
of finding optimal (lowest-cost) assignments.  By an {\em assignment}
we mean the set of choices of resource that are made by each task. For
the case of identical resources or identical tasks we obtain bounds on
the ratio between alternative Nash equilibria.

Our results show that the $L_1$ metric behaves very different to the
$L_\infty$ metric. In the case of the $L_\infty$ metric, there always
exists an optimal assignment that is also Nash, but the costs of
different Nash assignments can differ a lot.  In the case of the $L_1$
metric, the costs of any optimal assignment and the cost of the
minimum-cost Nash assignment can be arbitrarily far away from each
other, but in a lot of cases the costs of different Nash assignments
can differ only by a constant factor.

\subsection{The model}
\label{sec:model}

Here is the model from \cite{kp} (which is introduced in the context
of networks, as mentioned above).  We are given a set~$R$ of
$m$~resources with delays $d_1\leq \ldots \leq d_m$.  We are also
given a set~$T$ of~$n$ tasks with weights $w_1,\ldots,w_n$.  We assume
that $w_i \ge 1$ for all $i$, and we let $W = \sum_{i=1}^n w_i$ denote
the total task load.  Each task will select one resource.  Thus, an
{\it assignment\/} is a vector $A=(A_1,\ldots,A_n)$ which assigns the
$i$th task to resource $A_i\in R$.  (In the language of game theory,
an assignment associates each task with a ``pure
strategy''.\footnote{\cite{kp} also considers mixed strategies. See
  Section~\ref{sec:alt}.})  Let ${\assign}=\{1,\ldots,m\}^{n}$ denote
the set of all assignments.  The {\em load\/} of resource $\ell$ in
assignment~$A$ is defined to be
$$L(\ell,A) = d_{\ell} \sum_{i\in T: A_i=\ell} w_i.$$ The load of
task~$i$ in assignment~$A$ is $L(A_i,A)$.  Finally, the (social) {\it
  cost\/} of assignment~$A$ is given by
$$C(A) = \sum_{i\in T} L(A_i,A).$$

The notion of ``selfish behavior'' that we study comes from the
game-theoretic notion of a Nash equilibrium.  An assignment~$A$ is a
{\em Nash equilibrium} if and only if no task can lower its own load
by changing its choice of resource (keeping the rest of the assignment
fixed).  More formally, $A$ is said to be a {\it Nash\/} assignment
if, for every task~$i$ and every resource~$\ell$, we have $L(A_i,A)
\leq L(\ell,A')$, where the assignment~$A'$ is derived from~$A$ by
re-assigning task~$i$ to resource~$\ell$, and making no other
change. We let $\Nash(T,R)$ denote the set of all Nash assignments for
problem instance~$(T,R)$.  When the problem instance is clear from the
context, we refer to this as~$\Nash$. For a given problem instance, we
study the {\em coordination ratio\/} from {\cite{kp}} which is the
ratio between the cost of the highest-cost Nash assignment and the
cost of the lowest-cost assignment. That is
$$ \frac{\max_{N\in \Nash}C(N)}{ \min_{A\in\assign} C(A) }.$$ This
ratio measures the extent to which the social cost increases if we use
a worst-case Nash equilibrium rather than an optimal assignment.  We
also study the ratio between the lowest cost of a Nash assignment and
the lowest cost of an (arbitrary) assignment and also the ratio
between the lowest cost of a Nash assignment and the highest cost of a
Nash assignment.

Note that throughout the paper we study the average cost-per-task.
The reader should not confuse this with the average cost-per-resource.
The latter is trivial to optimize (it is achieved by assigning all
tasks to the link with the lowest delay) but it is not natural.

\subsection{Results}

\subsubsection*{Section~\ref{sec:coordratio}: Coordination Ratio in
  Terms of Task Weight Range} Theorem~\ref{thm:social} in
Section~\ref{sec:coordratio} bounds the coordination ratio in terms of
the range over which the task weights vary. In particular, suppose
that all task weights $w_i$ lie in the range $[1,w_{\max}]$.  Then
$$\frac{\max_{N\in\Nash} C(N)}{\min_{A\in{\assign}} C(A)} \leq
    4w_{\max}.$$

Several of our results focus on the special cases in which the
resource delays are identical (Section~\ref{sec:identical_resources})
or the task weights are identical
(Section~\ref{sec:identical_weights}).  The results are summarized as
follows.

\subsubsection*{Section~\ref{sec:identical_resources}: 
Resources with Identical Delays}

\begin{enumerate}\itemsep0ex
\item (Lemma~\ref{lem:big_nash})
For every $n$, there is a problem instance with $n$ tasks with
weights in the range $[1,n^2]$ for which
$$\frac {\min_{N\in\Nash} C(N)} {\min_{A\in{\assign}} C(A)} \geq
\frac{n}{5}.$$ Note that this is the ratio of the best Nash cost to
the optimal cost of an assignment, hence it gives a lower bound on the
coordination ratio that is proportional to $\sqrt{w_{\max}}$, where
$w_{\max}$ is the ratio of largest to smallest task weights.  This
lower bound should be contrasted with Theorem~\ref{thm:social} which
gives an upper bound that is proportional to $w_{\max}$.  These two
results show that it is variability of task weights, as opposed to
resource delays, that may lead to a big coordination ratio.
\item (Theorem~\ref{thm:nash_ratio})
{Nash assignments satisfy the following relation:
$$\frac{\max_{N\in\Nash} C(N)}
{\min_{N\in\Nash} C(N)}
    \leq 3.$$ }
\item (Lemma~\ref{lem:nash_ratio_lowerbound})
For every $\epsilon>0$, there is an instance satisfying
$$\frac {\max_{N\in\Nash} C(N)}{\min_{N\in\Nash} C(N)}
    \geq \frac{5}{3} (1-\epsilon).$$
The size of the problem instance depends upon~$\epsilon$.
\end{enumerate}

\subsubsection*{Section~\ref{sec:identical_weights}: Tasks with
  Identical Weights}

Theorem~\ref{thm:social} gives an upper bound of~$4$ for the
coordination ratio in the case of identical weights. We also have the
following results.
\begin{enumerate}
\item (Lemma~\ref{lem:opt_nash_uniformdiff})
For any $\epsilon>0$ there is a problem instance for which
$$\frac
{\min_{N\in\Nash} C(N)}
{\min_{A\in{\assign}} C(A)}
\geq \frac{4}{3} - \epsilon.$$
\item {(Theorem~\ref{thm:nash_ratios}) The lowest-cost and
  highest-cost Nash assignments satisfy:
$$\frac {\max_{N\in\Nash} C(N)} {\min_{N\in\Nash} C(N)} \leq
\frac{4}{3}$$ which is an exact result; we show that 
$4/3$ is obtainable for some
instance.} 

\item (Theorems~\ref{thm:findopt} and~\ref{thm:findoptnash}) {We give
  algorithms for finding a lowest-cost assignment and a lowest-cost
  Nash assignment.  These algorithms run in time $\bigo(mn)$.}
\end{enumerate}

\subsubsection*{Section~\ref{sec:dp}: Finding social optima using
  dynamic programming}
In Section~\ref{sec:dp} we show how dynamic
programming can be used to find optimal assignments under the $L_1$
metric, in either the identical-tasks case, or the identical-resources
case.  The algorithms extend to the case where either the task sizes
or the delays may take a limited set of values. This extension is used
to give approximation schemes for the cases where instead of a limit
on the number of distinct values, we have a limit on the ratio of
largest to smallest values.

\subsection{Alternative models and related work}
\label{sec:alt}

There are two collections of work related to our paper.  The first
uses a similar model, but a different cost function.  The second uses
a similar cost function, but a different model.

The model that we study was introduced by Koutsoupias and
Papadimitriou~\cite{kp}, who initiated the study of coordination
ratios.  They worked in the more general setting of {\em mixed
  strategies}.  In a mixed strategy, instead of choosing a
resource~$A_i$, task~$i$ chooses a vector $(p_{i,1},\ldots,p_{i,m})$
in which $p_{i,j}$ denotes the probability with which task~$i$ will
use resource~$j$. A collection of mixed strategies (one strategy for
each task) is a Nash equilibrium if no task can reduce its expected
cost by modifying its own probability vector.  Unlike us, Koutsoupias
and Papadimitriou measure social cost in terms of the $L_\infty$
metric.  Thus, the cost of a collection of strategies is the
(expected) maximum load of a resource (maximized over all resources).
Their coordination ratio is the ratio between the maximum cost
(maximized over all Nash equilibria) divided by the cost of the
optimal solution.  Koutsoupias and Papadimitriou give bounds on the
coordination ratio. These bounds are improved by Mavronicolas and
Spirakis~\cite{ms01}, and by Czumaj and V\" ocking~\cite{CV} who gave
an asymptotically tight bound. Fotakis et al.~\cite{fkkms02} consider
the same model.  They study the following algorithmic problems:
constructing a Nash equilibrium, constructing the worst Nash
equilibrium, and computing the cost of a given Nash equilibrium.  For
our purposes, we note that the existence of at least one pure Nash
assignment (as defined in Section~\ref{sec:model}) was also proven
in~\cite{fkkms02}.  Czumaj et al.~\cite{ckv02} give further results
for the model of~\cite{kp} using the $L_\infty$ metric for a wide
class of so-called {\em simple} cost functions.  They call a cost
function simple if it depends only on the injected load of the
resources.  They also show that for some families of simple monotone
cost functions, these results can be carried over to the $L_1$
metric. These are qualitative results relating the boundedness of the
coordination ratio in terms of boundedness of the bicriteria ratio. The 
bicriteria ratio describes by how many times the number of injected tasks 
must be decreased so that the worst case cost in a Nash equilibrium cannot 
exceed the optimal cost for the original tasks. In
contrast, here we are studying quantative bounds on the coordination
ratio for a special case of non-simple cost functions.

In \cite{glmms03} Gairing et al. study the combinatorial structure and
computational complexity of {\em extreme Nash equilibria},
i.e.\ equilibria that maximize or minimize the objective
function. Their results provide substantial evidence for the {\em
  Fully Mixed Nash Equilibrium Conjecture}, which states that the
worst case Nash equilibrium is the fully mixed Nash equilibrium where
each user chooses each link with positive probability.  They also
develop some algorithms for {\em Nashification}, which is the problem
of transforming an arbitrary pure strategy profile into a pure Nash
equilibrium without increasing the social cost. In \cite{fglmr03}
Feldmann et al.\ give a polynomial time algorithm for Nashification
and a polynomial time approximation scheme (PTAS) for computing a Nash
equilibrium with minimum social cost. In \cite{lmmrsv03} L\"ucking et
al.\ continue to study the Fully Mixed Nash Equilibrium Conjecture and
report substantial progress towards identifying the validity. Note
that all these publications use the $L_\infty$ metric to measure the
social cost.

Roughgarden and Tardos~\cite{RT} study coordination ratios in the
setting of traffic routing.  A problem instance specifies the rate of
traffic between each pair of nodes in an arbitrary network.  Each
agent controls a small fraction of the overall traffic.  Like us,
Roughgarden and Tardos use an $L_1$ cost-measure.  That is, the cost
of a routing is the sum of the costs of the agents.  The model of
Roughgarden and Tardos is in one sense much more general than our
model (from~\cite{kp}) which corresponds to a two-node network with
many parallel links.  However, most work in the model of~\cite{RT}
relies on the simplifying assumption that each agent can split its
traffic arbitrarily amongst different paths in the network.  In our
model, this assumption would correspond to allowing a task to split
itself between the resources, dividing its weight into arbitrary
proportions --- a simplification which would make our problems
trivial.  In particular, this simplification forces all Nash
assignments to have the same $L_1$ cost, which is not true in the
unsplittable model that we study.  In fact, in~\cite{RT} it is
demonstrated that if agents are not allowed to split their traffic
arbitrarily but each chooses a single path on which to route their own
traffic, then the cost of a Nash assignment can be arbitrarily larger
than an optimal (lowest-cost) assignment.  This is in contrast to
their elegant coordination ratio~\cite{RT} for the variant that they
study.  Even in our model, the splittable-task variant is useful as a
proof device.  In Section~\ref{sec:coordratio}, we use the
splittable-task setting to derive a lower bound on the cost of Nash
assignments in our model.  For other interesting results in the model
of Roughgarden and Tardos, see~\cite{RT} and~\cite{r03}.

Finally, we should contrast this work with~\cite{lmmr04} which (in the
model from~\cite{kp}) studies ``quadratic social cost'', a sum of
individual costs weighted by the task weights. That measure of social
cost is the same as ours in the case where all task weights are equal,
but in general leads to very different results for social optima and
coordination ratio, even in the special case of identical resources.

\section{Coordination Ratio in Terms of Task Weight Range}
\label{sec:coordratio}

Suppose that the weights lie in the range $[1,w_{\max}]$.
The purpose of this section is to prove Theorem~\ref{thm:social}, which 
shows that the coordination ratio is at most~$4w_{\max}$.

\begin{definition}\label{def:fa}
A {\em fractional assignment} $A^F$ for an instance $(T,R)$ is a
collection of real numbers $h_t(\ell)$ for $t\in T, \ell\in R$, such
that $0 \leq h_t(\ell) \leq 1$ and $\sum_{\ell\in R} h_t(\ell) = 1$ 
for all $t\in T$.  

If $A^F$ is a fractional assignment, the load of resource $\ell$
is defined as $L(\ell,A^F) = d_{\ell} \sum_{i\in T} w_i h_i(\ell)$.
The cost of task $i$ is defined as
$C_i(A^F) = \sum_{\ell\in R} h_i(\ell) L(\ell,A^F)$ and the cost of
$A^F$ is defined as $C(A^F) = \sum_{i\in T} C_i(A^F)$.

An {\em integral assignment} is a fractional assignment where all the
quantities $h_t(\ell)$ are equal to 0 or 1.  Note that we reserve the
notation $A$ (or $A(T,R)$ to denote the sets of tasks and resources)
strictly for integral assignments.

Define the {\em throughput} of resource set $R$ to be $D =
\sum_{\ell\in R} \frac{1}{d_\ell}$.
\end{definition}
We use Definition~\ref{def:fa} to provide a lower bound on the cost of
any integral assignment for a given instance $(T,R)$.  We start by
giving a lower bound on the cost of a fractional assignment.  The
following lemma is essentially the same as Lemma~2.5 of~\cite{RT}.

\begin{lemma}
\label{lem:optsplit}
If all tasks have weight $1$, then the optimal fractional assignment
$A^{F,opt}$ gives each resource a load of $n/D$ and therefore any task
$t$ has $C_t(A^{F,opt})=n/D$.
\end{lemma}

\begin{proof}
Let $x_\ell = \sum_{i\in T} h_i(\ell)$. From Definition~\ref{def:fa},
the load of resource $\ell$ is $x_\ell d_\ell$.
We have:
\begin{equation}\label{eqn:totalload}
\sum_{\ell \in R} x_{\ell} = n.
\end{equation}
Similar,
$$C(A^F) = \sum_{i\in T} C_i(A^F)
  =  \sum_{i\in T}  \sum_{\ell\in R} h_i(\ell) L(\ell,A^F)
  =  \sum_{i\in T}  \sum_{\ell\in R} h_i(\ell) d_\ell \sum_{j\in T} 
h_j(\ell)$$
where we have used $w_i=1$ in the expression for $L(\ell,A^F)$.  Thus,
$$C(A^F) =  \sum_{i\in T}  \sum_{\ell\in R} h_i(\ell) x_\ell d_\ell
=  \sum_{\ell\in R}  \sum_{i\in T} h_i(\ell) x_\ell d_\ell
=  \sum_{\ell\in R} x_\ell d_\ell \sum_{i\in T} h_i(\ell)
=  \sum_{\ell\in R} x_\ell^2 d_{\ell}.$$

Equation~(\ref{eqn:totalload}) gives a linear constraint on the
$x_\ell$ values, and we have expressed $C(A^F)$ in terms of the
$x_\ell$ values.  To minimise $C(A^F)$ subject
to~(\ref{eqn:totalload}) we use the well-known method of Lagrange 
multipliers (see \cite{MS}).  This means that the gradient of $C(A^F)$ 
and that of the
function $\sum_{\ell\in R} x_{\ell}$ must have the same direction:
$$\exists\Lambda\in\reals\textrm{   such that    } 
\nabla (C(A^F)) = \Lambda \nabla \Bigl( \sum_{\ell \in R} x_{\ell} \Bigr)$$
$${\rm i.e.}~~~~( 2d_1 x_1, 2 d_2 x_2, \ldots, 
 2 d_m x_m) =
(\Lambda,\Lambda,\ldots,\Lambda).$$

Hence, at the optimum we see that $x_{\ell} =
\frac{\Lambda}{2d_{\ell}}$ for all $\ell$.
Using~(\ref{eqn:totalload}), we then find that $x_\ell = \frac{n}{D
  d_{\ell}}$, and $L(\ell,A^{F,opt})=x_\ell d_{\ell}=n/D$ for all
$\ell\in R$.  Finally, for any task $i$
$$C_i(A^{F,opt}) = \sum_{\ell\in R} h_i(\ell) L(\ell,A^{F,opt})
= \sum_{\ell\in R} h_i(\ell) \frac{n}{D} = \frac{n}{D}\sum_{\ell\in R} 
h_i(\ell)
= \frac{n}{D}.$$
\end{proof}

The above result provides a useful lower bound on the cost of any
integral assignment $A$. We make one refinement for the lower bound:
note that if $m>n$, then any Nash or optimal assignment will only use
$n$ resources having smallest delays.\footnote{If the number of
  resources is allowed to be large by comparison with the number of
  tasks, then the optimal fractional assignment can be made
  artificially much lower than any integral assignment, by including a
  large number of resources with very large delays, thereby inflating
  the value of $D$.}  Hence an instance $(T,R)$ with $m>n$ can be
modified by removing the $m-n$ resources with largest delay.  In what
follows, we shall therefore make the assumption that $n\geq m$.  We
next proceed to give a bound on the coordination ratio for tasks
having weights in the range $[1,w_{max}]$.  We first give a definition
and an observation that will be useful to us.

\begin{definition}\label{def:fast}
Given a set $R$ of $m$ resources and a set of $n\geq m$ tasks, we say
resource $\ell$ is {\em fast} provided that $d_{\ell} \leq 2n/D$,
otherwise $\ell$ is {\em slow}.
\end{definition}

Given a set of tasks $T$, let $T^*$ denote a set of tasks such that
$|T^*| = |T|$ and each task $t\in T^*$ has unit weight.  We first make
an observation about the slow and fast resources for the optimal
fractional assignment $A^{F,opt}(T^*,R)$.

\begin{observation}\label{obs:mostfast}
For any sets $T,R$, in the optimal fractional assignment for the
instance $(T^*,R)$ 
we have 
$$\sum_{\ell\in R; \ell~{\rm fast}}\sum_{i\in T^*} h_i(\ell) \geq n/2.$$
\end{observation}

\begin{proof}
Let $A^{F,opt}$ denote an optimal fractional assignment.  
First note that $\sum_{\ell\in R}\sum_{i\in T^*} h_i(\ell) = n$.

Using Lemma~\ref{lem:optsplit} (and the definition of a ``slow resource'')
we find that in $A^{F,opt}(T^*,R)$ each 
slow resource $\ell$
satisfies $\sum_{i\in T^*} h_i(\ell) \leq 1/2$. Since we 
assume $n\geq m$, at most $n$ resources are slow, so that
$\sum_{\ell\in R; \ell~{\rm slow}}\sum_{i\in T^*} h_i(\ell) \leq n/2$.
The result follows from
$$\sum_{\ell\in R; \ell~{\rm fast}}\sum_{i\in T^*} h_i(\ell) =
\sum_{\ell\in R}\sum_{i\in T^*} h_i(\ell) -
\sum_{\ell\in R; \ell~{\rm slow}}\sum_{i\in T^*} h_i(\ell).$$
\end{proof}

\noindent
Here is our bound on the coordination ratio for tasks having weights
in the range $[1,w_{max}]$.  

\begin{theorem}\label{thm:social}
Suppose $(T,R)$ is a problem instance with $n$ tasks having weights in
the range $[1,w_{\max}]$ and $m$ resources. Then
$$ \max_{N\in\Nash} C(N) \leq  4w_{\max} \min_{A\in{\assign}} C(A).$$
\end{theorem}

\begin{proof}
Following our comments preceeding Definition~\ref{def:fast} we again
assume that $n\geq m$.  Let $\assign^F(T,R)$ denote the set of all
fractional assignments for the instance $(T,R)$.  As before, we let
$T^*$ denote the set of unit-weight tasks, where $|T^*|=|T|$.  We
first note that
\begin{equation}\label{ineq:lowerbound}
 \min_{A\in\assign(T,R)} C(A) \geq \min_{A^F\in\assign^F(T,R)} C(A^F)
      \geq \min_{A^F\in\assign^F(T^*,R)} C(A^F) = \frac{n^2}{D}.
\end{equation}
The last equality is an application of Lemma~\ref{lem:optsplit} to the
instance $(T^*,R)$.  We show that in any integral Nash assignment $N$,
all tasks $i$ satisfy the inequality $L(N_i,N) \leq 4w_{max} (n/D)$.
This would then imply that
$$\max_{N\in\Nash} C(N) = \max_{N\in\Nash}\sum_{i\in T} L(N_i,N) \leq
4w_{max}\left( \frac{n^2}{D} \right).$$
This, together with (\ref{ineq:lowerbound}), gives us the result.

Let $N$ denote a Nash assignment.  Suppose that under this assignment
some resource $j$ satisfies
$$ L(j,N) > 4w_{\max} \Bigl( \frac{n}{D} \Bigr). $$
We prove that $N$ is not Nash, by finding an assignment
$N'$ (obtained from $N$) by transferring one task 
from resource $j$ to some $j'$ such that
$$ L(j',N') \leq 4w_{\max} \Bigl( \frac{n}{D} \Bigr). $$
We start by proving there exists a fast 
resource $j'$ such that $L(j',N) \leq 2w_{\max}(\frac{n}{D}).$
To prove this, suppose for a contradiction that all fast
resources $\ell$ satisfy
\begin{equation}\label{eqn:contradict}
L(\ell,N) > 2w_{\max} \Bigl( \frac{n}{D} \Bigr).
\end{equation}
Let $A^{F,opt}$ denote an optimal fractional assignment for the
instance $(T^*,R)$.  We recall from Lemma~\ref{lem:optsplit}
that $L(\ell,A^{F,opt})=\frac{n}{D}$ for all resources $\ell$.  
Thus, if a fast resource $\ell$ satisfies~(\ref{eqn:contradict}),
we must have $L(\ell,N)/d_\ell > 2w_{\max} L(\ell,A^{F,opt})/d_\ell$.  
This means that
\begin{equation}\label{eqn:contradict2} 
  \sum_{i\in T;N_i = \ell} w_i >  2w_{\max} \sum_{i\in T^*} h_i(\ell) 
\end{equation}
where $h_i(\ell)$ are the values for the optimal fractional 
assignment $A^{F,opt}$.  
However, from Observation~\ref{obs:mostfast} we know that in $A^{F,opt}$
  $$\sum_{\ell\in R; \ell~{\rm fast}}\sum_{i\in T^*} h_i(\ell)
    \geq \frac{n}{2}$$
which, with Equation~(\ref{eqn:contradict2}) implies
$$\sum_{\ell\in R; \ell~{\rm fast}} \sum_{i\in T: N_i = \ell} w_i
     > \frac{n}{2}(2w_{\max}) = nw_{\max}.$$
This is a contradiction since the left hand side of this inequality 
(which is at most the sum of weights in the instance $(T,R)$) 
is at most $nw_{max}$.
Since we have a contradiction, we instead conclude 
there exists a fast resource $j'$ where
$$ L(j',N) \leq 2w_{\max} \Bigl( \frac{n}{D} \Bigr). $$

We now show how to construct $N'$ from $N$, thereby proving that $N$ 
was not a Nash assignment, a contradiction.
Recall since $j'$ is a fast resource, $ d_{j'} \leq \frac{2n}{D}$.
We consider two cases for $j'$. Let $k=L(j',N)/d_{j'}$.
If  $k\leq w_{\max}$, then moving one task from resource $j$ to 
resource $j'$ (to get the new assignment $N'$), we find that 
$$L(j',N') \leq d_{j'} (k+w_{\max})
    \leq 2\Bigl( \frac{n}{D} \Bigr) (w_{\max}+w_{\max})
    \leq 4w_{\max} \Bigl( \frac{n}{D} \Bigr).$$
If instead $k>w_{\max}$, then moving one task from $j$
to $j'$ to get $N'$, we find
$$ L(j',N') \le d_{j'} (k+w_{\max}) \leq d_{j'} \cdot 2k 
    = 2 L(j',N) \leq 4w_{\max} \Bigl( \frac{n}{D} \Bigr). $$
In either case, we have shown that $N$ is not a Nash assignment because
we can move one task (currently having a load greater 
than $4w_{max}(\frac{n}{D})$) from resource $j$ to resource $j'$ 
where it has a lower load.  
Thus, we conclude that if $N$ is a Nash assignment, then 
$L(j,N) \leq 4w_{max}\Bigl( \frac{n}{D} \Bigr)$ for all resources $j$, 
as desired to prove the theorem.  
\end{proof}


\section{Resources with Identical Delay}\label{sec:identical_resources}

In this section, we restrict our attention to problem instances with
identical delays, i.e.\ $d_1 = d_2 = \cdots = d_m$.  If we examine the
cost function we are using, we see that if all of the delays are
identical, we can factor this term from the cost.  Therefore, without
loss of generality, we can assume that for all~$i$, $d_i = 1$.

\paragraph{Notation:} Recall that $W=\sum_{t\in T} w_i$ denotes the total
weight of tasks. Let $\Lav$ be the average load on a resource, that
is, $\Lav =\frac{1}{m} \sum_{\ell\in R} L(\ell,A)=W/m$.  
Note in the case of identical (unit) delays, 
$\Lav$ is the same constant value for {\em all} assignments 
associated with a given problem instance $(T,R)$.

The following observation will be used in the proof of
Theorem~\ref{thm:nash_ratio}.

\begin{observation}\label{obs:lemn1}
Suppose $N\in \Nash$.  Every task~$i$ with $w_i > \Lav$ has its own
resource (which is not shared) in~$N$.
\end{observation}

\begin{proof}
Suppose to the contrary that task~$i$ shares a resource with task~$j$.
The load of task~$j$ is at least $w_j+ w_i$.  There must be some
resource whose load is at most the average load~$\Lav$, and task~$j$
would prefer to move to this resource, obtaining a new load of at most
$w_j+\Lav$.
\end{proof}

The next lemma shows that in the case of identical resources, the
ratio between the cost of the minimum (and, hence, any) Nash
assignment and the lowest cost of any assignment can be arbitrarily
large.  In fact, our example needs just two resources to obtain this
result.

\begin{lemma}\label{lem:big_nash}
For every $n>2$, there is an instance having identical resources,
and $n$ tasks with weights in the range $[1,n^2]$
for which the following holds:
$$\min_{N \in\Nash} C(N) \geq \frac{n}{5} \min_{A\in{\assign}} C(A).$$
\end{lemma}

\begin{proof}
For our problem instance we take
$m=2$, $d_1=d_2=1$, $w_1=w_2=n^2$, and $w_3=\cdots=w_n=1$.

Any assignment in which tasks~$1$ and~$2$ use the same resource is in
${\assign} - \Nash$ because one of these tasks could move to decrease its 
own load.  Thus, any $N\in \Nash$ will have tasks~$1$ and~$2$ on different
resources, which implies $C(N)\geq n^3$.  On the other hand,
$\min_{A\in{\assign}} C(A) \leq C(A^*)$, where $A^*$ is the assignment
which assigns tasks~$1$ and~$2$ to resource~$1$ and the other tasks to
resource~$2$.  $C(A^*)=4n^2+(n-2)(n-2)\leq 5n^2$.  Putting these facts
together, for every $N\in \Nash$,
$$C(N) \geq \frac{n}{5} \min_{A\in{\assign}} C(A).$$
\end{proof}

\noindent{\bf Remark:\quad} 
The example from the lemma has $w_{\max}=n^2$ and $w_{\min}=1$, 
showing that in this case
$C(N) \geq \frac{\sqrt{w_{\max}}}{5} \min_{A\in{\assign}} C(A)$.
Thus,  
the bound of Theorem~\ref{thm:social} needs to 
be some function of $w_{\max}$.
The example in Section 5.3 of~\cite{RT}
gives an observation similar to Lemma~\ref{lem:big_nash} for the
general-flow setting.  The example is a four-node problem instance
with two agents. The latency functions may be chosen so that there is
a Nash equilibrium which is arbitrarily worse than the social optimum.

Lemma~\ref{lem:big_nash} shows that the cost of the best
assignment and the cost of the best Nash assignment can be arbitrarily far
apart.
On the other hand, we can show that
the costs of different Nash assignments are close to one another.

\begin{theorem}\label{thm:nash_ratio}
For every instance with identical resources we have
$$\max_{N\in\Nash} C(N) \leq 3  \min_{N\in\Nash} C(N).$$
\end{theorem}

\begin{proof}
We first reduce the case in which $T$ contains a task with
$w_i>\Lav$ to the case in which $T$ does not contain such a task.
Let $(T',R')$ be a problem instance derived from $(T,R)$ by
removing a task~$i$ with $w_i>\Lav$ and removing one resource.
Then by Observation~\ref{obs:lemn1},
$$\max_{N\in\Nash(T,R)} C(N) = w_i + \max_{N\in\Nash(T',R')} C(N).$$
Similarly,
$$\min_{N\in\Nash(T,R)} C(N) = w_i + \min_{N\in\Nash(T',R')} C(N).$$
Thus, to prove the theorem, we only need to show
$$\max_{N\in\Nash(T,R)} C(N) \leq 3  \min_{N\in\Nash(T,R)} C(N)$$
for problem instances $(T,R)$ in which
every task has $w_i\leq \Lav$.
Let $(T,R)$ be such an instance.
Consider task $i$ having weight $w_i$. In a Nash assignment $A$, the load 
of task $i$ satisfies 
\begin{equation}\label{eq:id_nash_lower}
L(A_i,A) \geq \max \{ w_i, \Lav /2 \}
\end{equation}
since all resources must have load at least $\Lav/2$.
(If a resource has load less than $\Lav/2$ then
there must be a resource with load strictly larger than $\Lav$ 
with at least $2$ tasks on it, because of our assumption that 
$w_t\leq\Lav$ for all tasks $t$.  Then one of the tasks on this 
heavily loaded resource would move to the other less loaded one.)
Since $A$ is a Nash assignment, the load of task $i$ satisfies 
\begin{equation}\label{eq:id_nash_upper}
L(A_i,A) \leq \Lav + w_i.
\end{equation}

The ratio of the upper bound from (\ref{eq:id_nash_upper}) and
the lower bound from~(\ref{eq:id_nash_lower}) is at most $3$, attained when
$w_i = \Lav/2$. Hence the ratio between total costs (which is the ratio
between sums of individual task costs) is upper bounded by 3.
\end{proof}

The following lemma should be compared to Theorem~\ref{thm:nash_ratio}.

\begin{lemma}\label{lem:nash_ratio_lowerbound}
For every $\epsilon>0$, there is an instance with identical
resources such that
$$\min_{N\in\Nash} C(N) \leq
        \frac{3}{5} (1+\epsilon) \max_{N\in\Nash} C(N).$$
(The weights and number of tasks in this constructed instance
are allowed to depend upon~$\epsilon$.)
\end{lemma}

\begin{proof}
The number of tasks $n$ is equal to $6M+13$ where 
$M = \lceil \frac{2}{\epsilon} \rceil$.  $T$ will denote a 
set of tasks consisting of $6$ tasks of weight $3M$, 
$6$ tasks of weight $6M$, and $6M+1$ tasks of
weight $1$. In this case $R$ consists of $6$ resources.  
Let $N^{(1)}$ be the following Nash assignment:

$$\begin{tabular}{|c|c|c|} \hline
Resource   &  Tasks/Resource  & Cost/Resource \\  \hline \hline
$1$  &  $6M+1$ tasks, each of weight $1$ &  $6M+1$ \\ \hline
  $2,3,4$  &  $2$ tasks, each of weight $6M$ &  $12M$ \\ \hline
  $5,6$  &  $3$ tasks, each of weight $3M$ & $9M$ \\ \hline
\end{tabular}$$
Then $C(N^{(1)}) = (6M+1)\cdot (6M+1) + 6\cdot 12M  + 6 \cdot  9M
  = 36M^2 + 138M +1$.  
  Let $N^{(2)}$ be the following Nash assignment:

$$\begin{tabular}{|c|c|c|} \hline
Resource   &  Tasks/Resource  & Cost/Resource \\  \hline\hline
$1,2,3,4,5$  &  $1$ task of weight $6M$; $1$ task of weight $3M$; 
     $M$ tasks of weight $1$ &  $10M$ \\ \hline
  $6$  &    $1$ task of weight $6M$; $1$ task of weight $3M$; 
     $M+1$ tasks of weight $1$ &  $10M+1$ \\
\hline
\end{tabular}$$
In this case we have $C(N^{(2)}) \geq n\cdot 10M = (6M+13)10M$.
$$\frac{\min_{N\in\Nash} C(N)}{\max_{N\in\Nash} C(N)}
\leq \frac{C(N^{(1)})}{C(N^{(2)})}
\leq \frac{36M^2+138M+1}{10M(6M+13)}
\leq \frac{3}{5} \Bigl( 1 + \frac{11}{6M+13} \Bigr)
\leq \frac{3}{5} \Bigl( 1 + \frac{11}{\frac{12}{\epsilon}+13} \Bigr)
\leq \frac{3}{5} ( 1 + \epsilon)$$

\end{proof}


\section{Tasks with Identical Weights}\label{sec:identical_weights}

In this section, we turn our attention to instances in which the
weights of the tasks are identical, but the delays may be diverse.
Section~\ref{sec:weight_alg} is algorithmic in nature.  There, we
present an algorithm that constructs a lowest-cost assignment and an
algorithm for finding a Nash assignment with lowest possible cost.  In
Section~\ref{sec:weight_cost}, we compare the cost of Nash assignments
to the cost of the best-possible assignment and we compare the cost of
the best Nash assignment to the cost of the worst.  The comparisons
use structural observations arising from the algorithms in
Section~\ref{sec:weight_alg}.

\paragraph{Definitions:} Without loss of generality, we assume
that each task has unit weight.  Recall that $d_1 \leq d_2 \leq \cdots
\leq d_m$.  In this section, we use alternative notation to
represent an assignment.  In particular, an assignment will be denoted
as $\altA=\langle n_1,\ldots, n_m\rangle$, where $n_\ell$ is the {\em
number} of tasks assigned to resource $\ell$.  Thus $L(\ell,\altA) =
n_\ell d_\ell$ and $C(\altA) = \sum_\ell (n^2_\ell d_\ell)$.
Note that an assignment $\altA$ is a Nash assignment
if and only if $n_i d_i \leq (n_j+1) d_j$ for all $i$, $j$.

\subsection{Algorithmic Results}\label{sec:weight_alg}

We start with a structural observation about lowest-cost assignments.

\begin{lemma}\label{lem:add_one_task}
Suppose that $\overline{n}$ is a lowest-cost assignment for problem
instance~$(T,R)$. Let~$(T',R)$ be the problem instance derived
from~$(T,R)$ by adding one task.  Let $k$ be any resource that
minimizes the quantity $(2n_k + 1)d_k$.  Let $\overline{\psi}$ be the
assignment for~$(T',R)$ which agrees with~$\overline{n}$ except that
$\psi_k = n_k+1$.  Then $\overline{\psi}$ is a lowest-cost assignment
for~$(T',R)$.
\end{lemma}

\begin{proof}
We first argue that the problem instance $(T',R)$ has a lowest-cost
assignment~$\overline{\nu}$ with $\nu_k \geq \psi_k$.  To see this,
suppose that $\overline{\sigma}$ is a lowest-cost assignment
for~$(T',R)$ with $\sigma_k < \psi_k$.  Let $j$ be a resource with
$\sigma_j > \psi_j$.  Let $\overline{\nu}$ be the assignment
for~$(T',R)$ that agrees with~$\overline{\sigma}$ except that $\nu_k =
\sigma_k+1$ and $\nu_j = \sigma_j -1$.  Then
\begin{eqnarray}
C(\overline{\nu}) &=& C(\overline{\sigma}) +
({(\sigma_k+1)}^2 - \sigma_k^2)d_k + ({(\sigma_j-1)}^2 - 
\sigma_j^2)d_j\nonumber \\
  & =    & C(\overline{\sigma}) + (2 \sigma_k + 1)d_k - 
           (2 \sigma_j - 1)d_j\nonumber  \\
  & \leq & C(\overline{\sigma}) + (2 n_k + 1)d_k - 
           (2 \sigma_j - 1)d_j  \label{ineq:build1}\\
  & \leq & C(\overline{\sigma}) + (2 n_j + 1)d_j - 
           (2 \sigma_j - 1)d_j \label{ineq:build2} \\
  & \leq & C(\overline{\sigma}) + (2 n_j + 1)d_j  - 
           (2 n_j + 1)d_j \label{ineq:build3}\\
  &  =   & C(\overline{\sigma}),\nonumber
\end{eqnarray}
where~(\ref{ineq:build1}) follows from the upper bound on~$\sigma_k$,
(\ref{ineq:build2}) comes from the choice of~$k$,
and (\ref{ineq:build3}) comes from the choice of~$j$.  
So by iterating the above argument, we can take $\overline{\nu}$ 
to be a lowest-cost assignment for~$(T',R)$
satisfying $\nu_k\geq \psi_k$.

Suppose now that
$C(\overline{\nu})< C(\overline{\psi})$.
Let $\overline{y}$ be the assignment for $(T,R)$
that agrees with~$\nu$ on resources $\ell\neq k$ and has
$y_k = \nu_k-1$.
Then
\begin{eqnarray*}
C(\overline{\psi}) &=& C(\overline{n})+(\psi_k^2 - n_k^2)d_k
\leq C(\overline{y})+(\psi_k^2 - n_k^2)d_k\\
&=& C(\overline{y})+ (2 n_k + 1)d_k
\leq
C(\overline{y})+(2 \nu_k - 1)d_k =
C(\overline{y})+(\nu_k^2 - {(\nu_k-1)}^2)d_k = C(\overline{\nu}),
\end{eqnarray*}
where the first inequality follows from the optimality
of~$\overline{n}$,
giving a contradiction to our assumption on the costs 
of $\overline{\nu}$ and $\overline{\psi}$.  Therefore 
$\overline{\psi}$ is a lowest-cost assignment for $(T',R)$.
\end{proof}

\noindent
Theorem~\ref{thm:findopt} follows directly from
Lemma~\ref{lem:add_one_task}.

\begin{figure}[ht]
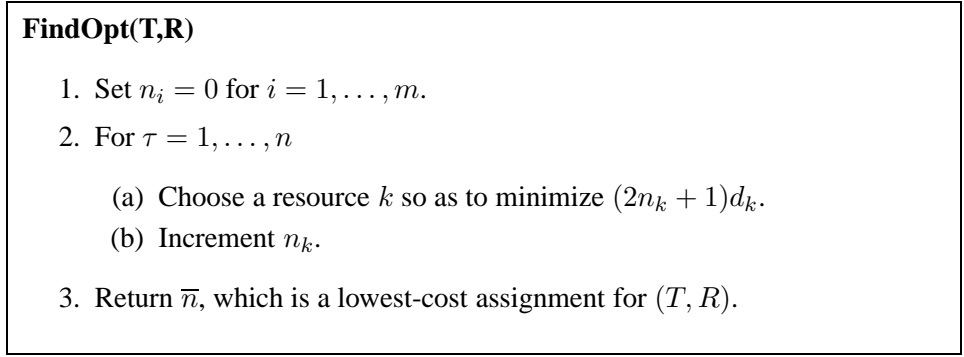

\begin{center}
\fbox{
\begin{minipage}{\boxwidth}
\smallskip
{\bf FindOpt(T,R)}
\begin{enumerate}\itemsep0ex
\item Set $n_i = 0$ for $i = 1,\ldots,m$.
\item For $\tau = 1,\ldots,n$
     \begin{enumerate}\itemsep0ex
     \item Choose a resource $k$ so as to minimize $(2n_k+1)d_k$.
     \item Increment $n_k$.
     \end{enumerate}
\item Return $\overline{n}$, which is a lowest-cost assignment for
$(T,R)$.
\end{enumerate}
\smallskip
\end{minipage}
}
\caption{An algorithm for constructing a lowest-cost assignment
for a problem instance $(T,R)$ with $n\geq 1$ tasks and $m$ resources.}
\label{fig:findopt}
\end{center}
\end{figure}

\begin{theorem}
Let $(T,R)$ be a problem instance with $n\geq 1$ tasks and
$m$~resources.
Algorithm {\bf FindOpt} (see Figure~\ref{fig:findopt}) constructs
a lowest-cost assignment for~$(T,R)$ in $\bigo(n m)$ time.
\label{thm:findopt}
\end{theorem}

If $n=\Omega(m)$ then the algorithm can be sped up to $O(n \log m)$ by using, 
for example, a heap to store the queue of resources. A similar improvement can 
be made to algorithm {\bf FindOptNash}, which follows.
The following lemmas give information about the structure of
Nash assignments.

\begin{lemma}\label{lem:different_nash}
If $\overline{\nu}\in \Nash(T,R)$
and $\overline{\rho}\in \Nash(T,R)$ then, for any $j\in R$,
$|\nu_j - \rho_j|\leq 1$.
\end{lemma}

\begin{proof}
Suppose $\rho_\ell > \nu_\ell$. Let $k$ be a resource such
that $\rho_k < \nu_k$.
Then since $\overline{\rho}$ and $\overline{\nu}$ are Nash assignments,
$ \rho_\ell d_\ell \leq (\rho_k+1) d_k\leq \nu_k d_k \leq
(\nu_\ell+1) d_\ell$,
so $\rho_\ell \leq \nu_\ell+1$.
\end{proof}

\begin{lemma}
Suppose $\overline{n}\in \Nash(T,R)$.
If $n_i > n_j$ then $d_i \leq d_j$.
\label{lemswitch}
\end{lemma}
\begin{proof}
Suppose to the contrary that $n_i>n_j$ and $d_i > d_j$.
Then $(n_j+1)d_j < n_i d_i$, so $\overline{n}$ is not a
Nash assignment.
\end{proof}

\begin{figure}[ht]
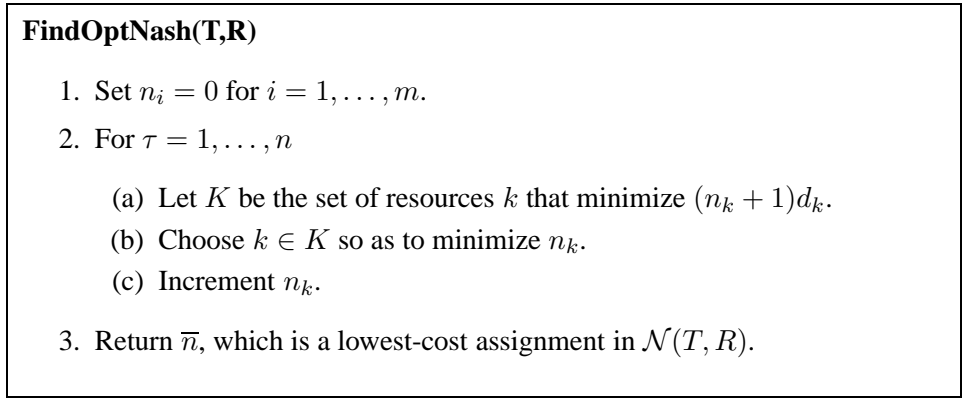

\begin{center}
\fbox{
\begin{minipage}{\boxwidth}
\smallskip
{\bf FindOptNash(T,R)}
\begin{enumerate}\itemsep0ex
\item Set $n_i = 0$ for $i = 1,\ldots,m$.
\item For $\tau = 1,\ldots,n$
     \begin{enumerate}\itemsep0ex
     \item Let $K$ be the set of resources~$k$ that 
minimize~$(n_k+1) d_k$.
     \item Choose $k\in K$ so as to minimize $n_k$.
     \item Increment $n_k$.
     \end{enumerate}
\item Return $\overline{n}$, which is a lowest-cost assignment in
$\Nash(T,R)$.
\end{enumerate}
\smallskip
\end{minipage}
}
\caption{An algorithm for constructing a lowest-cost Nash assignment
for a problem instance $(T,R)$ with $n\geq 1$ tasks and $m$ resources.}
\label{fig:findoptnash}
\end{center}
\end{figure}

\begin{theorem}
Let $(T,R)$ be a problem instance with $n\geq 1$ tasks and
$m$~resources.
Algorithm {\bf FindOptNash} (see Figure~\ref{fig:findoptnash})
constructs
a lowest-cost assignment in~$\Nash(T,R)$ in $\bigo(n m)$ time.
\label{thm:findoptnash}
\end{theorem}

\begin{proof}
First note that the algorithm maintains the invariant that the
assignment for tasks $1,\ldots,j$ on resources in~$R$ is a Nash
assignment. This follows from the fact that $k$~is chosen so as to
minimize $(n_k+1) d_k$. We prove by induction on~$n$ that the
constructed assignment has lowest cost amongst Nash assignments. The
base case is $n=1$. For the inductive step, let $\overline{n}$ be the
(optimal) Nash assignment for a problem instance $(T,R)$ with
$n$~tasks constructed by the algorithm.  
Derive~$(T',R)$ from~$(T,R)$ by adding one task. Let
$\overline{\nu}$ be the assignment constructed by {\bf
FindOptNash(T',R)}. Let $i$ be the resource such that $\nu_i = n_i +
1$. Suppose for contradiction that $\overline{\rho}\in \Nash(T',R)$
satisfies $C(\overline{\rho})< C(\overline{\nu})$. By
Lemma~\ref{lem:different_nash}, there are three cases.

\noindent{\bf Case 1:\quad} $\rho_i = \nu_i = n_i+1$\newline
Since $C(\overline{\rho})< C(\overline{\nu})$
there are resources~$j$ and~$\ell$ in~$R$ such that
$\rho_j = \nu_j - 1$ and $\rho_\ell = \nu_\ell+1$
and
\begin{equation}
\rho_j^2 d_j + \rho_\ell^2 d_\ell < \nu_j^2 d_j + \nu_\ell^2 d_\ell =
n_j^2 d_j + n_\ell^2 d_\ell.
\label{eq:nashopt1}
\end{equation}
Let $\overline{\psi}$ be the assignment constructed by the algorithm
just before the $n_j$th task is assigned to resource~$j$.
Then
\begin{equation}
(\psi_\ell+1) d_\ell
\leq (n_\ell+1) d_\ell
= \rho_\ell d_\ell
\leq (\rho_j+1) d_j
= n_j d_j
= (\psi_j+1) d_j,
\label{eq:leslie}
\end{equation}
where the second inequality follows from the fact that
$\overline{\rho}$ is a Nash assignment.  Because the algorithm chose
resource~$j$ rather than resource~$\ell$, all of the inequalities in
Equation~(\ref{eq:leslie}) are equalities so
\begin{equation}
n_j d_j = (n_\ell+1) d_\ell.
\label{zzone}
\end{equation}
Furthermore, by step 2b of the algorithm, $\psi_j \leq \psi_\ell$
so
$n_j-1 = \psi_j \leq \psi_\ell \leq n_\ell$ which, together
with Equation~(\ref{zzone})
implies
\begin{equation}
\label{zztwo}
d_j\geq d_\ell.
\end{equation}
Finally, the following calculation contradicts
Equation~(\ref{eq:nashopt1}).
\begin{eqnarray*}
\rho_j^2 d_j + \rho_\ell^2 d_\ell
&=& {(n_j-1)}^2 d_j + {(n_\ell+1)}^2 d_\ell\\
&=& n_j^2 d_j + n_\ell^2 d_\ell + (2 n_\ell + 1) d_\ell - (2 n_j - 
1) d_j\\
&=& n_j^2 d_j + n_\ell^2 d_\ell +2(n_\ell + 1) d_\ell-(2 n_j - 
1) d_j- d_\ell\\
&=& n_j^2 d_j + n_\ell^2 d_\ell +2 n_j d_j- (2 n_j - 1) d_j- d_\ell\\
&\geq& n_j^2 d_j + n_\ell^2 d_\ell.
\end{eqnarray*}
The final equality follows from~(\ref{zzone}) and the
inequality follows from~(\ref{zztwo}).

\noindent{\bf Case 2:\quad} $\rho_i = \nu_i-1 = n_i$\newline
We will construct an assignment $\overline{\sigma}\in \Nash(T',R)$ with
$C(\overline{\sigma}) \leq C(\overline{\rho})$
and $\sigma_i = \nu_i$. Case~$1$ then applies to~$\overline{\sigma}$.
Let $j$ be a resource with $\rho_j> \nu_j$, so by
Lemma~\ref{lem:different_nash} $\rho_j = \nu_j+1$.
Since $\overline{\nu}$ is a Nash assignment,
\begin{equation}\label{eq:nashopt4}
(n_i+1) d_i = \nu_i d_i \leq (\nu_j+1) d_j = (n_j+1) d_j.
\end{equation}
Since $\overline{\rho}$ is a Nash assignment,
\begin{equation}
\label{eq:nashopt6}
(n_j+1) d_j = \rho_j d_j \leq (\rho_i+1) d_i = \nu_i d_i = (n_i+1) d_i.
\end{equation}
Inequalities~(\ref{eq:nashopt4}) and~(\ref{eq:nashopt6})
together imply
\begin{equation}\label{eq:nashopt7}
(n_i+1) d_i = (n_j+1) d_j
\end{equation}
and
\begin{equation}
\label{eq:nashopt8}
(\rho_i+1) d_i = \rho_j d_j.
\end{equation}
Since the algorithm chose to assign the last task in~$(T',R)$
to resource~$i$ rather to resource~$j$ (in step 2b),
we have $n_i\leq n_j$.
Lemma~\ref{lemswitch} and
Equation~(\ref{eq:nashopt7})
imply that $d_i\geq d_j$.

Let $\overline{\sigma}$ be the assignment that agrees
with~$\overline{\rho}$
except $\sigma_i = \rho_i+1$ and $\sigma_j = \rho_j - 1$.
Equation~(\ref{eq:nashopt8})
implies the following facts since $\overline{\rho}$ is a Nash
assignment.
\begin{enumerate}
\item for $\ell \not\in\{i,j\}$,
$(\rho_i+1) d_i = \rho_j d_j \leq (\rho_\ell+1) d_\ell$,
\item for $\ell \not\in\{i,j\}$,
$\rho_j d_j = (\rho_i+1) d_i \geq \rho_\ell d_\ell$.
\end{enumerate}
The first of these implies that
$\sigma_i d_i \leq (\sigma_\ell+1) d_\ell$
and the second implies that
$(\sigma_j+1) d_j \geq \sigma_\ell d_\ell$ for all $\ell$.
Thus, $\sigma$ is a Nash assignment.
The argument that
$C(\overline{\sigma}) \leq C(\overline{\rho})$
is exactly the same as the end of Case~1.
\begin{eqnarray*}
C(\overline{\sigma}) - C(\overline{\rho}) &=&
(\sigma_i^2 - \rho_i^2) d_i + (\sigma_j^2 - \rho_j^2) d_j\\
&=& (2 \rho_i+1) d_i - (2 \rho_j-1) d_j\\
&=& 2 \rho_j d_j - d_i - (2 \rho_j-1) d_j\\
&=& -d_i + d_j\\
&\leq& 0,
\end{eqnarray*}
where the second-to-last equality uses Equation~(\ref{eq:nashopt8}).

\noindent{\bf Case 3:\quad} $\rho_i = \nu_i+1 = n_i+2$\newline
As in Case~2, we
construct an assignment $\overline{\sigma}\in \Nash(T',R)$ with
$C(\overline{\sigma}) \leq C(\overline{\rho})$
and $\sigma_i = \nu_i$.
The argument is similar to Case~2, but is included for completeness.
Let $j$ be a resource with $\rho_j < \nu_j$, so by
Lemma~\ref{lem:different_nash} $\rho_j = \nu_j-1$.
Since $\overline{\nu}$ is a Nash assignment,
\begin{equation}\label{eq:newnashopt4}
n_j d_j = \nu_j d_j \leq (\nu_i + 1) d_i = (n_i + 2) d_i.
\end{equation}
Since $\overline{\rho}$ is a Nash assignment,
\begin{equation}
\label{eq:newnashopt6}
(n_i+2) d_i = \rho_i d_i \leq (\rho_j+1) d_j = n_j d_j.
\end{equation}
Inequalities~(\ref{eq:newnashopt4}) and~(\ref{eq:newnashopt6})
together imply
\begin{equation}\label{eq:newnashopt7}
(n_i+2) d_i = \rho_i d_i = (\rho_j+1) d_j = n_j d_j.
\end{equation}

Let $\overline{\sigma}$ be the assignment that agrees
with~$\overline{\rho}$
except $\sigma_i = \rho_i-1$ and $\sigma_j = \rho_j + 1$.
Equation~(\ref{eq:newnashopt7})
implies the following facts since $\overline{\rho}$ is a Nash
assignment.
\begin{enumerate}
\item for $\ell \not\in\{i,j\}$,
$(\rho_j+1) d_j = \rho_i d_i \leq (\rho_\ell+1) d_\ell$,
\item for $\ell \not\in\{i,j\}$,
$\rho_i d_i = (\rho_j+1) d_j \geq \rho_\ell d_\ell$.
\end{enumerate}
The first of these implies that
$\sigma_j d_j \leq (\sigma_\ell+1) d_\ell$
and the second implies that
$(\sigma_i+1) d_i \geq \sigma_\ell d_\ell$.
Thus, $\sigma$ is a Nash assignment.
Finally,
\begin{eqnarray*}
C(\overline{\sigma}) - C(\overline{\rho})
&=& (\sigma_i^2 - \rho_i^2) d_i + (\sigma_j^2 - \rho_j^2) d_j   \\
&=& ({(n_i+1)}^2 - {(n_i+2)}^2) d_i + (n_j^2 - {(n_j-1)}^2) d_j \\
&=& (2 n_j- 1) d_j - (2 n_i+3) d_i \\
&=& n_j d_j + (n_j- 1) d_j - (n_i+2) d_i - (n_i+1) d_i \\
&=& (n_j-1) d_j - (n_i+1) d_i \\
&\leq& n_j d_j -  (n_i+1) d_i\\
&\leq& 0,
\end{eqnarray*}
since $\overline{n}$ is a Nash assignment.
Note that we use Equation~(\ref{eq:newnashopt7}) in the last equality.

\null{} From the three cases together we see that the algorithm
{\bf FindOptNash} indeed finds an optimal Nash assignment.
\end{proof}

\subsection{Comparison of Optimal and Nash Costs}\label{sec:weight_cost}

Our first result shows that even for identical tasks the minimum cost of
a Nash assignment can be larger then the optimal cost.

\begin{lemma}\label{lem:opt_nash_uniformdiff}
With identical task weights, for all $\epsilon>0$ there is an instance
for which $$\min_{N\in\Nash} C(N) \geq
  \left(\frac{4}{3}-\epsilon\right) \min_{A\in{\assign}} C(A).$$
\end{lemma}

\begin{proof}
Consider the instance
with $m=2$, $d_1=1/2$, $d_2=(1+\epsilon)$, $n=2$, $w_1=1$ and $w_2=1$.
There are three  assignments. The assignment $\overline{n}=\langle 2,0
\rangle$
has $L(1,\overline{n})=1$ and $C(\overline{n})=2$.  This assignment is
a Nash
assignment, because moving one of the tasks to resource~$2$ would give
it a new load of~$1+\epsilon$.  The assignment $\overline{\rho}=\langle
1,1\rangle$
has $L(1,\overline{\rho})=(1/2)$, $L(2,\overline{\rho})=1+\epsilon$ and
$C(\overline{\rho})=1.5+\epsilon$.
This assignment is not a Nash assignment, because the task on resource
$2$
could move to resource~$1$ for a new load of~$1$.  Finally,
the assignment $\overline{\psi}=\langle 0,2\rangle$ has
$L(2,\overline{\psi})=2(1+\epsilon)$. It
is not a Nash assignment, because either task could move to resource~$1$
for a new load of~$1/2$.  Thus, $\overline{n}$ is the only member of
$\Nash$ and 
$$C(\overline{n})\geq\left(\frac{2}{1.5+\epsilon}\right)\min_{A\in{\assign}}
C(A) \geq  \left(\frac{4}{3}-\epsilon\right) \min_{A\in{\assign}} C(A).$$
\end{proof}

\noindent
In the example from the proof of 
Lemma~\ref{lem:opt_nash_uniformdiff}
there is only one Nash assignment, and its cost
is almost $4/3$ times the cost of the best assignment. If we do the
same construction with $\epsilon=0$,
we obtain an instance with two different Nash equilibria that differ in cost
from each other by a factor $4/3$. 
The following theorem shows that $4/3$ is in fact
the largest ratio obtainable between alternative Nash equilibria
for any problem instance where task weights are identical.

\begin{theorem}\label{thm:nash_ratios}
Suppose the tasks weights are identical. For the ratio between
the lowest-cost Nash assignment and the highest-cost Nash assignments we have
$$ \max_{N\in\Nash} C(N)
  \leq \frac{4}{3} \min_{N\in\Nash} C(N).$$
\end{theorem}

\begin{proof}
Suppose that $\overline{n}$ and $\overline{\rho}$ are distinct
assignments in $\Nash(T,R)$.
Suppose that $\ell$ is a resource for which $n_\ell>\rho_\ell$.
By Lemma~\ref{lem:different_nash},
$n_\ell = \rho_\ell+1$. Also, there is a resource $\ell'$
for which $n_{\ell'}<\rho_{\ell'}$. Again, by Lemma~\ref{lem:different_nash},
$n_{\ell'}+1=\rho_{\ell'}$.
We will show that
\begin{equation}
\label{eqgoal}
\rho_\ell^2 d_\ell + \rho_{\ell'}^2 d_{\ell'} \leq \frac43
\left(
n_\ell^2 d_\ell + n_{\ell'}^2 d_{\ell'}
\right),
\end{equation}
which proves the theorem since the resources on which~$\overline{n}$
and~$\overline{\rho}$ differ can be partitioned into pairs
such as the pair~$\ell,\ell'$. Now 
\begin{equation}
\label{eqmain}
\rho_\ell^2 d_\ell + \rho_{\ell'}^2 d_{\ell'} =
{(n_\ell-1)}^2 d_\ell + {(n_{\ell'}+1)}^2 d_{\ell'}.
\end{equation}
Since $\overline{n}$ is a Nash assignment,
$d_\ell n_\ell \leq d_{\ell'}(n_{\ell'}+1) = d_{\ell'} \rho_{\ell'}$
and since $\overline{\rho}$ is a Nash assignment,
$d_{\ell'} \rho_{\ell'} \leq d_\ell (\rho_\ell+1) = d_\ell n_\ell$
so $d_\ell n_\ell = d_{\ell'}\rho_{\ell'}$.
Now
if $n_{\ell'}=0$ then the right-hand side of~(\ref{eqmain})
is
$${(n_\ell-1)}^2 d_\ell + {(n_{\ell'}+1)} d_{\ell'}
={(n_\ell-1)}^2 d_\ell + {\rho_{\ell'}} d_{\ell'}
={(n_\ell-1)}^2 d_\ell + {n_{\ell}} d_{\ell}
\leq n_{\ell}^2 d_\ell,
$$
so~(\ref{eqgoal}) holds.
So suppose that $n_{\ell'}\geq 1$.
Note that, for any $A\geq 1$,
the right-hand side of (\ref{eqmain})
is at most
$$
A \left({(n_\ell-1)}n_\ell d_\ell + \frac{n_{\ell'}+1}{A}
{(n_{\ell'}+1)} d_{\ell'}
\right).
$$
We will choose
$A = ({n_{\ell'}^2+ 2 n_{\ell'} + 1})/
({n_{\ell'}^2+n_{\ell'}+1})$
so
$({n_{\ell'}+1})/{A} = 1 + 
{n_{\ell'}^2}/({n_{\ell'}+1})$.
Plugging this in, we get that the right-hand side of (\ref{eqmain})
is 
at most
$$
A \left({(n_\ell-1)}n_\ell d_\ell + 
{(n_{\ell'}+1)} d_{\ell'}
+ n_{\ell'}^2 d_{\ell'}
\right)
=
A \left({(n_\ell-1)}n_\ell d_\ell + 
{n_{\ell}} d_{\ell}
+ n_{\ell'}^2 d_{\ell'}
\right)
=
A \left(n^2_\ell d_\ell + n_{\ell'}^2 d_{\ell'}
\right)
.$$
Equation~(\ref{eqgoal})
follows from the observation that $A\leq4/3$ 
for every $n_{\ell'}\geq 1$.
\end{proof}

\section{Finding Optima with Dynamic Programming}\label{sec:dp}

In~\cite{fkkms02}, the authors present a polynomial time greedy
algorithm for computing a Nash assignment for the $L_\infty$ cost
function. The algorithm works as follows. It considers each of the
tasks in the order of non-increasing weights and assigns them to the
resource that minimized their delay.

In this last section we give dynamic programming algorithms that find
minimum-cost assignments for the various special cases that we have
studied. These algorithms extend from the identical tasks (respectively,
identical resources) case to the case where there are $O(1)$ distinct
values that may be taken by the task weights (respectively, resource
delays). The algorithms extend to give approximation schemes for the
case where there is a $O(1)$ bound on the ratio between the largest
and smallest task weights (respectively, largest to smallest delays),
as studied in Theorem~\ref{thm:social}.

\begin{lemma}
\label{tasks_ordered}
There exists an optimal assignment in which the set $R$ of resources
can be ordered in such a way that if $i\in R$ precedes $j\in R$, then
all tasks assigned to $i$ have weight less than or equal to all tasks
assigned to $j$.
\end{lemma}

\begin{proof}
Suppose that we have an assignment $A$ where the resources cannot
be ordered in this way. Then there exist two resources $i$ and
$j$, with two tasks assigned to $i$ having weights $w$ and
$w'$, and a task assigned to $j$ with weight $w''$, such that
$w < w'' < w'$.
Let $n_i$ and $n_j$ be the numbers of tasks assigned to $i$ and
$j$ respectively, and let $d_i$ and $d_j$ be their delays.
Let $W_i = L(i,A)/d_i$ and $W_j = L(j,A)/d_j$. The total cost of
tasks assigned to $i$ and $j$ is $W_i n_i d_i + W_j n_j d_j$.

In the following we consider 3 cases.
If $n_i d_i > n_j d_j$ then we may exchange the tasks with
weights $w''$ and $w'$ to reduce the social cost $C(A)$ (the operation
reduces $W_i$ by $w'-w''$ and increases $W_j$ by $w'-w''$).
If $n_i d_i < n_j d_j$ then we may exchange the tasks with weights
$w$ and $w''$ to reduce $C(A)$. In both cases $A$ is suboptimal.

If $n_i d_i = n_j d_j$ we may make either exchange since they both
leave the social cost unchanged. In the following we build up the
order iteratively and assume that all occurrences of case 1 and case 2
are already eliminated.  Suppose we have any optimal assignment and
that some subset of the resources have been placed in order, say $R_1
<= R_2 <= \cdots <= R_c$.  Consider adding another of the resources to
the order that we are constructing.  Perhaps the new resource, $R$, is
greater than the ordered resources $R_1,\leq,R_{a-1}$ but it cannot be
placed either below or above resource $R_a$.  This is because $R_a$
has tasks $w$ and $w'$ and resource $R$ has task $w''$ with $w<w''<w'$
as above.  Since the assignment is optimal, we are in the case $n_i
d_i = n_j d_j$ from above, and we can exchange $w''$ with $w'$. This
leaves the order of the original subset, $R_1,\ldots,R_c$, unchanged.
We continue this process until $R$ has bigger tasks, and then we can
continue adding it to the order.

\end{proof}

\begin{theorem}\label{thm5.2}
Suppose that $m$ resources have unit delay. Then an optimal
assignment of $n$ tasks with arbitrary weights to those resources
may be found in time $O(n^2m)$.
\end{theorem}

\begin{proof}
We may order the task weights so that $w_1\geq w_2 \geq \ldots w_n$.
Let $C_{j,k}$ be the cost of an optimal assignment of tasks with
weights $w_1,\ldots,w_j$ to resources $r_1,\ldots,r_k$. We want
to compute the quantity $C_{n,m}$.

Lemma~\ref{tasks_ordered} guarantees an optimal assignment of the
tasks to a set of resources that will assign the $\ell$ lowest-weight
tasks to some resource, for some value of $\ell$.  $C_{j,k}$ may be
found by, for each $\ell \in \{1,2,\ldots,j\}$, assign tasks with
weights $w_{j+1-\ell},\ldots,w_j$ to resource $r_k$.
$$C_{j,k} = \min_{\ell\in\{0,1,2,\ldots,j\} } \Bigl(
C_{j-\ell,k-1} + \ell\cdot (w_{j+1-\ell}+\ldots+w_j) \Bigr).$$

$C_{n,m}$ can be found using a dynamic programming table of size
$O(nm)$ each of whose entries is computed in time $O(n)$.
\end{proof}

The above dynamic program extends to the case where delays may belong
to a set of $O(1)$ elements $\{d_1,\ldots,d_\alpha\}$ where $\alpha$
is a constant. Let $m_\ell$ be the number of resources with delay
$d_\ell$, so that $m=m_1+\ldots+m_\alpha$.

Let $C_{j,k_1,k_2,\ldots k_\alpha}$ be the cost of an optimal
assignment of tasks with weights $w_1,\ldots,w_j$ to a set of
resources containing $k_\ell$ resources with delay $d_\ell$, for
$\ell=1,2,\ldots,\alpha$.  Lemma~\ref{tasks_ordered} guarantees an
optimal assignment that will (for some $\ell$ and $\ell'$) assign the
$\ell$ lowest weight tasks to some resource with delay $d_{\ell'}$,
provided $k_{\ell'}>0$.

$$C_{j,k_1,k_2,\ldots k_\alpha} =
\min_{\ell\in\{0,1,2,\ldots,j\}; \ell'\in\{1,2,\ldots,\alpha\}
  \mbox{with $k_{\ell'}>0$} } \Bigl(
C_{j-\ell,k_1,k_2,\ldots,k_{\ell'}-1,\ldots,k_\alpha} +
\ell \cdot d_{\ell'} \cdot (w_{j+1-\ell}+\ldots+w_j) \Bigr).$$
The dynamic programming table has size $O(nm^\alpha)$ and
each entry is computed in time $O(n)$.

\bigskip
The following theorem generalises the algorithm {\bf FindOpt} to the
case where there is an $O(1)$ bound on the number of distinct values
taken by task weights.

\begin{theorem}
Let weights $w_1,\ldots,w_n$ take values in $\{w'_1,\ldots,w'_\alpha\}$.
Let $n_\ell$ be the number of tasks with weight $w'_\ell$, so that
$n=n_1+\ldots+n_\alpha$.
Given delays $d_1\leq d_2\leq\ldots d_m$, we may find an optimal
assignment in time $O(mn^{2\alpha})$.
\end{theorem}

\begin{proof}
Let $C_{k,j_1,\ldots,j_\alpha}$ be the cost of an optimal assignment
to resources with delays $d_1,\ldots,d_k$ of a set of tasks
containing $j_\ell$ tasks of weight $w'_\ell$, for 
$\ell=1,2,\ldots,\alpha$.
For $x\in {\bf N}$, let $[x]$ denote the set $\{0,1,2,\ldots,x\}$.

$$C_{k,j_1,j_2,\ldots, j_\alpha} =
\min_{j'_1\in [j_1];j'_2\in [j_2];\ldots;
j'_\alpha\in [j_\alpha]} \Bigl(
C_{k-1,j_1-j'_1,\ldots,j_\alpha-j'_\alpha} +
(j'_1+\ldots+j'_\alpha)\cdot d_k \cdot
(w'_1\cdot j'_1+\ldots+w'_\alpha\cdot j'_\alpha) \Bigr).$$
There are $O(mn^\alpha)$ entries in the dynamic programming table, and
each entry is computed in time $O(n^\alpha)$.
\end{proof}

The above algorithm can be used to obtain an approximation scheme for
the case where there is a bound on the ratio of maximum to minimum
weights, as studied in Theorem~\ref{thm:social}.  Assume the weights
are indexed in non-ascending order, $w_1\geq w_2\geq\ldots \geq w_n$
and the ratio $w_1/w_n$ is upper-bounded by some pre-set limit
$\alpha$.

Let $\epsilon\le 1$ be the desired accuracy. Choose $k$ such that
$(w_1/w_n)^{1/k} \leq 1+\epsilon$. Take each weight and round it up to
the nearest value of $w_n\cdot (w_1/w_n)^{t/k}$ where $t$ is as small
as possible in $\{0,\ldots,k\}$. The new weights take $k+1$ distinct
values. An optimal assignment for the new weights has cost at most
$1+\epsilon$ times the cost of an optimal assignment for the old
weights, since each weight has increased by at most a factor
$1+\epsilon$. In this special case of fixed ratio of largest to
smallest task weight, $k$ depends only on $\epsilon$, and the
resulting algorithm has run time $O(mn^{2k})$ where $k=O(\epsilon^{-1}
\ln(w_1/w_n))$.

The dynamic programming algorithm of Theorem \ref{thm5.2} can be used
in exactly the same way to obtain an approximation scheme subject to a
fixed limit on the ratio of largest to smallest delay. The details are
omitted.

\section{Conclusions}

This paper studies a very general resource allocation problem.
We are given
a collection of resources  each of which is associated with a
``delay'' and a collection of tasks, each given with a weight.
We assume that each task chooses its resource in a selfish manner,
minimizing its own cost, and we are interested in
determining the social cost of this selfish behavior.  Previous work
on this problem has measured ``social cost'' in terms of the
$L_\infty$ metric -- that is, the longest delay incurred by any task.
Our measure of social cost is the $L_1$ metric -- that is, the average
delay (over tasks).

We give bounds on the size of the coordination ratio; we also study 
the algorithmic problem
of finding optimal (lowest-cost) assignments.   For
the case of identical resources or identical tasks we obtain bounds on
the ratio between alternative Nash equilibria.

Our results show that the $L_1$ metric behaves very differently to the
$L_\infty$ metric. In the case of the $L_\infty$ metric, there always
exists an optimal assignment that is also Nash, but the costs of
different Nash assignments can differ a lot.  In the case of the $L_1$
metric, the costs of any optimal assignment and the cost of the
minimum-cost Nash assignment can be arbitrarily far away from each
other, but in a lot of cases the costs of different Nash assignments
can differ only by a constant factor.

\end{document}